\documentclass[a4paper,11pt]{article}
\pdfoutput=1
\usepackage{tikz}
\usetikzlibrary{shapes.symbols, positioning, decorations.pathmorphing}
\usepackage{amsmath, array, amssymb}
\usepackage{dcolumn}
\newcolumntype{d}[1]{D{.}{.}{#1}}
\usepackage{mdframed}
\newmdenv[
  linecolor=black,
  linewidth=1pt,
  roundcorner=5pt,
  innertopmargin=10pt,
  innerbottommargin=10pt,
  innerrightmargin=10pt,
  innerleftmargin=10pt
]{keyresult}
\usepackage{JHEP/jheppub}          

\arxivnumber{2603.09858}

\title{\boldmath Polarization transfer in $\psi'\to\psi\pi\pi$: a complete spin density matrix analysis framework}

\author[a,b]{Jiabao Gong}
\author[a,b]{Guanyu Wang}
\author[c]{Dongyu Yuan}
\author[d]{Libo Liao}
\author[e]{Yilun Wang}
\author[f]{Jiarong Li}
\author[c]{Xiaoshen Kang}
\author[e]{Lei Zhang}
\author[d]{Jin Zhang}
\author[a,b,g]{Gang Li}

\affiliation[a]{Institute of High Energy Physics, Chinese Academy of Sciences, 100049, Beijing, China}
\affiliation[b]{University of Chinese Academy of Sciences, 100049, Beijing, China}
\affiliation[c]{Liaoning University, 110036, Liaoning, China}
\affiliation[d]{Sun Yat-sen University (Shenzhen), 518107, Guangdong, China}
\affiliation[e]{Nanjing University, 210093, Jiangsu, China}
\affiliation[f]{Tsinghua University, 100084, Beijing, China}
\affiliation[g]{Center for High Energy Physics, Henan Academy of Sciences,450000, Henan, China}
\emailAdd{gongjb@ihep.ac.cn}
\emailAdd{ligang@ihep.ac.cn}

\abstract{
A theoretical framework based on the Spin Density Matrix (SDM) formalism is developed to describe polarization transfer in the decay chain $e^+e^- \rightarrow \psi^\prime \rightarrow \psi\pi\pi$. Explicit relations connecting the SDMs of $\psi^\prime$ and $\psi$ are derived, generalizing Cahn's analysis into a complete SDM treatment. For the dominant $S$-wave $\pi\pi$ emission, the SDM is shown to be perfectly preserved, $\rho_\psi = \rho_{\psi^\prime}$, rendering the $\psi$ an ideal probe of the initial polarization state. Deviations arising from $D$-wave contributions are quantified, and a self-consistency experimental test is proposed that simultaneously validates the framework and constrains partial wave amplitudes. 
This formalism provides a consistent basis for extracting $\psi$ polarization and for amplitude analyses of subsequent $\psi$ decays in a continuum-background-free environment.
The framework extends to other hadronic transitions, including $\psi' \to h_c\pi^0$ in charmonium and $\Upsilon(nS) \to \Upsilon(mS)\pi\pi$ in bottomonium, as well as to electroweak processes such as $e^+e^- \to Z^\ast \to ZH$, where the same angular-momentum structure governs polarization transfer --- offering a unified probe of dynamics from charmonium to the Higgs sector.
}


\begin{document}
\maketitle
\flushbottom

\section{Introduction}

Polarization observables~\cite{Faccioli:2010kd,Faccioli:2010ej} in heavy quarkonium systems provide sensitive probes into both perturbative and non-perturbative aspects of Quantum Chromodynamics (QCD)\cite{Brambilla:2010cs,LHCb:2013izl}. With the increasing statistical precision achieved at the BESIII~\cite{BESIII:2009fln}, the Belle II~\cite{Belle-II:2010dht}, and the LHC experiments~\cite{Erhan:1995pk,CMS:1992tji,ATLAS:1994vge} enter the precision era, systematic effects rather than statistical limitations now dominate the overall uncertainty budget. These effects include background contamination, model dependence in amplitude analyses, and incomplete control of polarization correlations in decay chains.

The hadronic transition $\psi^\prime \rightarrow J/\psi\pi\pi$~\cite{Abrams:1975zp} (hereafter denoting $J/\psi$ as $\psi$ for brevity) provides a uniquely clean laboratory. With its large branching fraction~\cite{ParticleDataGroup:2024cfk} and the absence of continuum background, this decay provides a clean source of $\psi$ mesons, allowing its properties to be studied with minimal systematic contamination. Moreover, since the parent $\psi^\prime$ is produced with a well-defined polarization state in $e^+e^-$ annihilation, the $\psi^\prime \rightarrow J/\psi\pi\pi$ transition offers a unique laboratory to investigate how polarization is transmitted through a non-perturbative hadronic transition.

Beyond its cleanliness, this channel provides $\psi$ mesons with well-defined polarization characteristics. In $e^+e^-$ annihilation, the $\psi^\prime$ is produced predominantly with transverse polarization ($J_z = \pm1$). A fundamental question then arises: how does this polarization transfer to the daughter $\psi$? This transfer depends on the partial wave structure of the $\pi\pi$ system. 
It will be shown that in the limit of dominant $S$-wave emission~\cite{Brown:1975dz,Cahn:1975ts,Pham:1975zq,Voloshin2006,BES:1999guu,BESIII:2025ozb} 
the daughter SDM coincides with that of the parent after normalization,  $\rho_\psi = \rho_{\psi^\prime}$. 
This result follows from angular-momentum conservation and from the structure of the leading double electric-dipole ($E1E1$) transition in the QCD multipole expansion~\cite{Kuang:2012wp,Kuang:2006me}. 
Sub-leading partial waves induce controlled deviations that can be expressed in terms of measurable amplitudes.

The SDM formulation developed below serves three purposes. First, it establishes $\psi^\prime \rightarrow \psi\pi\pi$ as an ideal source of polarized $\psi$ mesons with precisely known SDM --- essential for amplitude analyses. Second, any deviation from $\rho_\psi = \rho_{\psi^\prime}$ directly quantifies $D$-wave contributions, providing a clean probe of sub-leading QCD dynamics. Third, the framework enables a powerful self-consistency test: the $\psi$ SDM calculated from the measured $\psi^\prime$ polarization and extracted $\pi\pi$ partial waves~\cite{BES:1999guu} can be compared with the directly measured $\psi$ SDM from its leptonic decay, validating the entire chain of reasoning.

Building on the foundational work in Refs.~\cite{Brown:1975dz,Cahn:1975ts}, the explicit transformation $\rho_\psi = T\rho_{\psi^\prime}T^\dagger$ is derived, where the transition matrix $T$ is expressed in terms of partial wave amplitudes. The analysis proves analytically that pure $S$-wave decay leads to exact polarization inheritance and quantifies modifications from $D$-wave contributions. Concrete experimental strategies are then outlined, including precision extraction of polarization parameters from dilepton angular distributions and a novel three-path consistency test.

The paper is organized as follows. Section 2 establishes notation and reviews the theoretical framework. Section 3 derives the polarization transfer relations, presenting the key identity for $S$-wave dominance and $D$-wave corrections. Section 4 discusses experimental applications, including measurement strategies and the self-consistent validation test. Section 5 extends the formalism to the related process $\psi^\prime \rightarrow h_c\pi^0$. Section 6 concludes with a discussion and conclusion. Explicit forms of the transition matrices are provided in Appendix A.

\section{Notations and theoretical framework}

\subsection{The spin density matrix formalism}

The spin density matrix completely describes the polarization state of any particle with arbitrary spin. For a vector meson such as $\psi^{(\prime)}$, the SDM in the helicity basis is a $3\times 3$ Hermitian matrix:
\begin{equation}\label{eq:sdm-definition}
\rho_{mm^\prime}, \qquad m,m^\prime \in \{+1,0,-1\},
\end{equation}
where $m$ labels the helicity state. The diagonal elements $\rho_{mm}$ represent the population probabilities of the corresponding helicity states, while off-diagonal elements $\rho_{mm^\prime}$ ($m \neq m^\prime$) encode quantum coherence arising from interference between different helicity amplitudes.

A central property of the SDM formalism is that matrix elements transform linearly under decay processes. For a decay $A \rightarrow B + X$, if the decay amplitude can be written as $\mathcal{M}_{b,a} = \langle b | a \rangle$ connecting initial state $|a\rangle$ of particle $A$ to final state $|b\rangle$ of particle $B$, then the SDM of $B$ is obtained via:
\begin{equation}
\rho^{(B)}_{bb^\prime} = \sum_{aa^\prime} \mathcal{M}_{ba} \, \rho^{(A)}_{aa^\prime} \, \mathcal{M}^*_{b^\prime a^\prime}.
\end{equation}
This transformation law allows precise tracking of polarization through decay chains, which is precisely what will be exploited for $\psi^\prime \rightarrow \psi\pi\pi$.

\subsection{Kinematic variables and quantum numbers}

\begin{figure}[htbp]
\centering 
\begin{tikzpicture}[
    line width=0.9pt, 
    font=\small, 
    scale=1.0 
]
    \draw (-3, 0.0) -- (3, 0.0);
    \draw (-3,-0.5) -- (3,-0.5);
    
    \node[below] at (-2,-0.6) {$c$};
    \node[above] at (-2, 0.0) {$\bar{c}$};
    \node[below] at ( 2,-0.6) {$c$};
    \node[above] at ( 2, 0.0) {$\bar{c}$};
    
    \draw[decorate, decoration={coil, segment length=2mm, radius=1mm}, black!70] 
         (-0.2, 0.20) -- (-0.2,1.30);
    \draw[decorate, decoration={coil, segment length=2mm, radius=1mm}, black!70] 
         ( 0.2, 0.20) -- ( 0.2,1.30);

    \fill (0, 1.75) circle (0.50);
    \fill (0,-0.25) circle (0.50);
    
    \draw[dashed] (0,2.0) -- (3,2.0) node[right] {$\pi$};
    \draw[dashed] (0,1.5) -- (3,1.5) node[right] {$\pi$};
    \node[right] at (3.5,1.75) {$\ell=0,2$};
    
    \node[left] at (-3.0,-0.2) {$\psi^\prime$: $m=\pm1$};
    \node[right] at (3.0,-0.2) {$\psi$: $s_z=\pm1$};
    \node[right] at (3.5,0.85) {$L=0,2,4$};
\end{tikzpicture}
\caption{Schematic diagram for the transition $\psi^\prime \to \psi\pi\pi$. The quantum numbers $(\ell, L)$ denote the orbital angular momentum of the $\pi\pi$ system and the relative motion between $(\pi\pi)$ and $\psi$, respectively. Parity and charge conjugation conservation restrict $\ell$ and $L$ to even values, with $\ell=0,2$ and $L=0,2$ being dominant. Blobs represent non-perturbative hadronization processes; wavy lines denote soft gluon exchanges mediating the transition.\label{fig:psipipi-diagram}}
\end{figure}
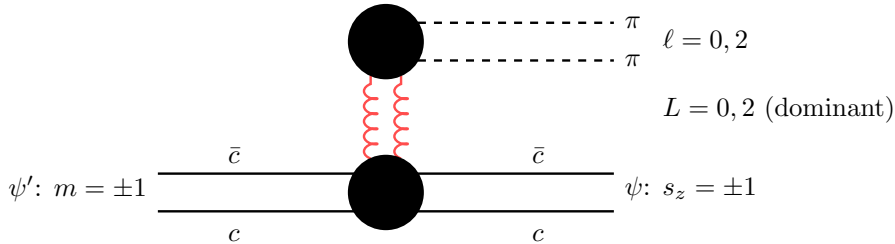

Consider the process:
\begin{equation}\label{eq:main-process}
e^+e^- \rightarrow \psi^\prime \rightarrow \psi\pi\pi, \quad \psi\to \ell^+\ell^-,
\end{equation}
illustrated schematically in Fig.~\ref{fig:psipipi-diagram}. 

Following Cahn's notation~\cite{Cahn:1975ts}, angular variables are defined in successive rest frames through Lorentz transformations. The solid angle $\Omega_\psi = (\theta_\psi, \phi_\psi)$ describes the direction of $\psi$ in the $\psi^\prime$ rest frame, while $\Omega_\pi = (\theta_\pi, \phi_\pi)$ specifies the direction of $\pi^+$ in the $\pi\pi$ rest frame. Finally, $\Omega_\ell = (\theta_\ell, \phi_\ell)$ denotes the direction of $\ell^+$ in the $\psi$ rest frame. These coordinate systems are obtained through successive Lorentz boosts, ensuring relativistic consistency despite the non-relativistic appearance of the angular momentum coupling formalism~\cite{Cahn:1975ts}.

The transition is characterized by several angular momentum quantum numbers. The vector $\vec{\ell}$ represents the orbital angular momentum of the $\pi\pi$ system with $z$-component $\ell_z$, while $\vec{s}$ and $\vec{J}$ denote the spins of $\psi$ and $\psi^\prime$ (both spin-1) with $z$-components $s_z$ and $m$ respectively. The orbital angular momentum between $\psi$ and $\pi\pi$ is designated as $\vec{L}$ with $z$-component $L_z$, and the channel spin $\vec{S} = \vec{s} + \vec{\ell}$ has $z$-component $S_z$. The partial wave amplitudes $M_{\ell LS}$ must satisfy conservation laws imposed by parity and charge conjugation symmetries. Parity conservation requires $L + \ell$ to be even, while charge conjugation invariance restricts $\ell$ to even values. Consequently, the dominant contributions arise from $M_{001}$ (pure $S$-wave with $\ell=0, L=0, S=1$), $M_{201}$ ($D$-wave $\pi\pi$ emission with $\ell=2, L=0, S=1$), and $M_{021}$ ($D$-wave relative motion with $\ell=0, L=2, S=1$)~\cite{BES:1999guu}.

\section{\texorpdfstring
  {Polarization transfer from $\psi^\prime$ to $\psi$}
  {Polarization transfer in psi' -> psi}
}
\subsection{General transformation law}

The amplitude for $\psi^\prime \rightarrow \psi\pi\pi$ in the helicity formalism is~\cite{Cahn:1975ts}:
\begin{equation}\label{eqn:amp-full}
T_{s_z, m}(\Omega_\psi, \Omega_\pi) = \sum_{\substack{\ell,L,S \\ \ell_z,L_z,S_z}} M_{\ell LS} 
Y_L^{L_z}(\Omega_\psi) Y_\ell^{\ell_z}(\Omega_\pi) 
C^{S;\ell,s}_{S_z;\ell_z s_z} C^{J;LS}_{m;L_z S_z},
\end{equation}
where $Y_J^M(\Omega)$ are spherical harmonics and $C^{j_1;j_2j_3}_{m_1;m_2m_3}$ are Clebsch-Gordan coefficients. This amplitude connects the initial $\psi^\prime$ helicity state $m$ to the final $\psi$ helicity state $s_z$.

The differential cross section for the full process in Eq.~\eqref{eq:main-process} factorizes as:
\begin{equation}\label{amp-factorize}
\frac{d\sigma}{d\Omega_\psi d\Omega_\pi d\Omega_\ell } \propto 
\sum_{\lambda,\lambda^\prime}\sum_{s_z,s_z^\prime} \sum_{m,m^\prime} D^{1*}_{s_z,\lambda}(\Omega_\ell)
T_{s_z m} \rho^\prime_{m,m^\prime}  T^*_{s^\prime_z m^\prime} 
D^{1}_{s^\prime_z,\lambda^\prime}(\Omega_\ell),
\end{equation}
where $D^1_{mm^\prime}(\Omega)$ are Wigner rotation matrices describing the $\psi \rightarrow \ell^+\ell^-$ decay, and $\lambda = \lambda_{\ell^+} - \lambda_{\ell^-} = \pm 1$ is the lepton helicity difference (restricted by QED to $\pm 1$).

From Eq.~\eqref{amp-factorize}, the transformation of the $\psi$ SDM is:
\begin{equation} \label{eqn:full-sdm-jpsi}
\rho_{s_z,s_z^{\prime}} =  \sum_{m,m^{\prime}}T_{s_z,m} \rho^{\prime}_{m,m^{\prime}} T_{s_z^{\prime},m^{\prime}}^\dagger,
\end{equation}
or in matrix notation:
\begin{equation}\label{eqn:full-sdm-jpsi-matrix}
{\rho = T\rho^\prime T^\dagger}
\end{equation}
Equation~\eqref{eqn:full-sdm-jpsi} or ~\eqref{eqn:full-sdm-jpsi-matrix} constitutes the fundamental polarization-transfer relation of this work. 
While Cahn~\cite{Cahn:1975ts} first analyzed the angular distributions of $\psi^\prime\to\psi\pi\pi$, the present work provides a complete SDM formulation of polarization transfer for this process.

Equation~\eqref{eqn:full-sdm-jpsi} or~\eqref{eqn:full-sdm-jpsi-matrix} defines a polarization-transfer relation that is independent of the production mechanism of the parent $\psi^\prime$ meson. The transition matrix $T$ is determined solely by the internal dynamics of the hadronic transition $\psi^\prime \to \psi\pi\pi$, whereas the initial spin density matrix $\rho^\prime$ fully encodes the production environment. As a result, the same transformation applies without modification to vector mesons produced in $e^+e^-$ annihilation, hadron--hadron collisions, photoproduction, or decays of heavier resonances. Moreover, the formalism is directly applicable to analogous transitions in other quarkonium systems, such as $\Upsilon(nS)\to\Upsilon(mS)\pi\pi$~\cite{BaBar:2008xay,Belle:2007xek}, making the polarization-transfer relations broadly relevant across heavy quarkonium spectroscopy. This universality is essential for transferring polarization information across experimental settings and energy scales.

\subsection{Partial wave decomposition}

To make the transformation explicit, $T$ is expanded in partial waves. Retaining only the dominant $S$- and $D$-wave contributions:
\begin{equation}\label{amp-part}
T_{s_z,m} = T^{001}_{s_z,m} + T^{201}_{s_z,m} + T^{021}_{s_z,m} + \mathcal{O}(\text{higher-partial-waves}),
\end{equation}
where:
\begin{align}
T^{001}_{s_z,m} &= M_{001} \delta_{m,s_z}, \\
T^{201}_{s_z,m} &= M_{201} \sum_{\ell_z, S_z} Y_\ell^{\ell_z}(\Omega_\pi) C^{S;\ell s}_{S_z;\ell_z s_z} \delta_{m,S_z}, \\
T^{021}_{s_z,m} &= M_{021} \sum_{L_z, S_z} Y_L^{L_z}(\Omega_\psi) C^{J;LS}_{m;L_z S_z}\delta_{s_z,S_z}.
\end{align}
Explicit matrix forms of these operators are provided in Appendix~\ref{sec:TlLS}.

\subsection{\texorpdfstring
  {The $S$-wave identity: perfect polarization transfer}
  {The S-wave identity: perfect polarization transfer}
}
For pure $S$-wave $\pi\pi$ emission ($\ell=0, L=0$), only $T^{001}$ contributes, and Eq.~\eqref{amp-part} reduces to:
\begin{equation}
T_{s_z,m} = M_{001} \delta_{s_z,m}.
\end{equation}
Substituting into Eq.~\eqref{eqn:full-sdm-jpsi}:
\begin{equation}
\rho_{s_z,s_z^\prime} = \sum_{m,m^\prime} M_{001} \delta_{s_z,m} \, \rho^\prime_{m,m^\prime} \, M_{001}^* \delta_{s_z^\prime,m^\prime}
= |M_{001}|^2 \rho^\prime_{s_z,s_z^\prime}.
\end{equation}
Since the transition matrix $T$ is not unitary in general, the daughter SDM is renormalized after integration over phase space to ensure $\mathrm{Tr}\,\rho=1$. After normalization, the fundamental result is obtained:

\textbf{Proposition ($S$-wave polarization inheritance):}
For pure $S$-wave $\pi\pi$ emission in $\psi^\prime \rightarrow \psi\pi\pi$, the spin density matrix is perfectly preserved:
\begin{equation}\label{eqn:sdm-swave-identity}
{\rho_\psi = \rho_{\psi^\prime}}
\end{equation}

\textit{Physical interpretation:} This identity follows directly from angular momentum conservation. When the $\pi\pi$ system carries zero orbital angular momentum ($\ell=0$) and is emitted in a relative $S$-wave with the $\psi$ ($L=0$), the decay amplitude factorizes as $\mathcal{M} \propto \vec{\epsilon}_{\psi^\prime} \cdot \vec{\epsilon}_\psi^*$, where $\vec{\epsilon}$ denotes polarization vectors. This scalar product structure prevents rotational mixing between helicity states: $m = s_z$ for all non-zero amplitudes. Consequently, the $\psi$ inherits the full polarization structure of its parent.
\textit{Equivalently,} the decay amplitude for pure $S$-wave emission is proportional to the scalar product of polarization vectors, $\mathcal{M} \propto \vec{\epsilon}_{\psi^\prime} \cdot \vec{\epsilon}_\psi^*$. This scalar product is invariant under rotations, preventing any helicity flip between the parent and daughter mesons. So, each event preserves the helicity quantum number ($m = s_z$), and after normalization one obtains $\rho_\psi = \rho_{\psi^\prime}$. 

\textit{QCD interpretation:} 
The exact preservation of the spin density matrix in the pure $S$-wave limit is not purely kinematic but follows from the structure of the leading $E1E1$ transition in the QCD multipole expansion. In the framework of the QCD multipole expansion, the $\psi^\prime \rightarrow \psi\pi\pi$ transition is dominated by the emission of two color-electric dipole ($E1E1$) gluons~\cite{Kuang:2012wp,Kuang:2006me,Yan:1980uh}, which couple to the heavy-quark spin only through the scalar product of polarization vectors. As a consequence, no helicity mixing occurs at leading order, and the daughter $\psi$ inherits the full polarization structure of the parent $\psi^\prime$.

Corrections to this identity arise from sub-leading multipole contributions and higher partial waves, such as $D$-wave $\pi\pi$ emission or relative motion, providing a sensitive probe of non-leading QCD dynamics.

\textit{Experimental significance:} Equation~\eqref{eqn:sdm-swave-identity} establishes $\psi^\prime \rightarrow \psi\pi\pi$ as an ideal source of polarized $\psi$ mesons with precisely known SDM. Deviations from this identity directly quantify non-
$S$-wave contributions.

\subsection{\texorpdfstring
  {Corrections from $D$-wave contributions}
  {Corrections from D-wave contributions}
}
While the $\pi\pi$ system is strongly $S$-wave dominated, small $D$-wave components have been observed~\cite{BES:1999guu,BESIII:2025ozb}. These introduce helicity-state mixing. Including $D$-wave terms, the transformation becomes:
\begin{equation}
\rho = T^{001}\rho^\prime T^{001\dagger} + T^{201}\rho^\prime T^{201\dagger} + T^{021}\rho^\prime T^{021\dagger},
\end{equation}
where cross terms $T^{001}\rho^\prime T^{201\dagger}$, $T^{001}\rho^\prime T^{021\dagger}$, and $T^{201}\rho^\prime T^{021\dagger}$ vanish upon angular integration because of orthogonality of spherical harmonics.

For a diagonal parent SDM $\rho^\prime = \text{diag}(p_+, 0, p_-)$ (appropriate for transversely polarized $\psi^\prime$ production), the diagonal elements of $\rho$ are given by:
\begin{align}
\rho_{+,+} &= |M_{001}|^2 p_+ + \frac{1}{10}\left[|M_{201}|^2 + |M_{021}|^2\right](p_+ + 6p_-), \label{eq:rhojpsi-diag1} \\
\rho_{0,0} &= \frac{3}{10}\left[|M_{201}|^2 + |M_{021}|^2\right](p_+ + p_-), \label{eq:rhojpsi-diag2} \\
\rho_{-,-} &= |M_{001}|^2 p_- + \frac{1}{10}\left[|M_{201}|^2 + |M_{021}|^2\right](6p_+ + p_-). \label{eq:rhojpsi-diag3}
\end{align}

These expressions demonstrate that $D$-wave contributions act as \textit{polarization mixers}, transferring population between helicity states. Even a modest $D$-wave fraction $f_D = (|M_{201}|^2 + |M_{021}|^2)/( |M_{001}|^2 + |M_{201}|^2 +|M_{021}|^2)$ can induce measurable modifications. For example, with $f_D = 0.03$~\cite{BES:1999guu} and pure transverse initial polarization ($\text{diag}(0.5, 0, 0.5)$), one obtains $\rho_{0,0} \approx 0.01$ instead of zero --- a numerically non-negligible effect at current experimental precision that can be observed with high-statistics datasets. 
However, this deviation remains negligibly small in partial wave analysis (PWA), since PWA fits are typically insensitive to polarization effects at the percent level. The smallness of this effect therefore justifies adopting $\text{diag}(0.5, 0, 0.5)$ as a valid approximation for the $\psi$ SDM.

\section{Experimental applications}

\subsection{Extraction of SDM from dilepton angular distributions}

The spin density matrix elements of both $\psi^\prime$ and $\psi$ can be extracted experimentally from the angular distributions of their dilepton decays. In the helicity frame adopted here, the decay distribution can be parametrized in terms of the polarization coefficients $(\lambda_\theta, \lambda_\phi, \lambda_{\theta\phi})$, which are directly related to the underlying SDM elements.

The processes $\psi^\prime \rightarrow \mu^+\mu^-$ (measuring $\rho^\prime$) and $\psi \rightarrow \mu^+\mu^-$ (measuring $\rho$) both yield angular distributions of the form~\cite{Faccioli:2010kd,Faccioli:2010ej}:
\begin{equation}\label{eq:WthetaPhi}
W(\theta,\phi) \propto 1 + \lambda_\theta \cos^2\theta + \lambda_\phi \sin^2\theta \cos 2\phi + \lambda_{\theta\phi}\sin 2\theta \cos\phi,
\end{equation}
where the polarization parameters are related to SDM elements by:
\begin{equation}\label{eqn:lambda-rho-connection}
\lambda_\theta = \frac{\rho_{+,+} + \rho_{-,-} - 2\rho_{0,0}}{\rho_{+,+} + \rho_{-,-} + 2\rho_{0,0}}, \quad
\lambda_\phi = \frac{2\text{Re}(\rho_{+,-})}{\rho_{+,+} + \rho_{-,-} + 2\rho_{0,0}}, \quad
\lambda_{\theta\phi} = \frac{\sqrt{2}\text{Re}(\rho_{+,0}-\rho_{0,-})}{\rho_{+,+} + \rho_{-,-} + 2\rho_{0,0}}.
\end{equation}

For pure $S$-wave and transverse polarization only ($\rho_{0,0} = 0$, $\rho_{+,-} \neq 0$), one obtains:
\begin{equation}
\lambda_\theta = 1, \qquad \lambda_\phi = 2\text{Re}(\rho_{+,-}), \qquad \lambda_{\theta\phi} = 0.
\end{equation}
In the helicity frame and neglecting higher-order QED corrections, the parameter $\lambda_\phi$ is directly related to transverse beam polarization through the relation $\lambda_\phi = P_T^2$ for beams with transverse polarization $P_T$.
\noindent
The assumption of purely transverse $\psi^\prime$ polarization ($\rho'_{00}=0$) can be experimentally verified from the same dataset using the decay $\psi'\to\mu^+\mu^-$. Any deviation from $\lambda_\theta=1$ in the $\psi'$ angular distribution would indicate a non-zero $\rho'_{00}$ due to initial-state radiation, beam energy spread, or detector acceptance asymmetries. Such a deviation must then be incorporated into the transformation $\rho = T\rho' T^\dagger$ as an input, and the self-consistency test of Fig.~\ref{fig:self-consistent-test} naturally cross-checks the measured $\rho'$ against the daughter $\psi$ SDM.

When $D$-wave contributions are present, combining Eqs.~\eqref{eq:rhojpsi-diag1}--\eqref{eq:rhojpsi-diag3} with Eq.~\eqref{eqn:lambda-rho-connection} yields:
\begin{equation}
\lambda_\theta = \frac{|M_{001}|^2 + \frac{1}{10}(|M_{201}|^2 + |M_{021}|^2)}{|M_{001}|^2 + \frac{13}{10}(|M_{201}|^2 + |M_{021}|^2)},
\end{equation}
representing a deviation from unity. For $f_D = 0.03$, this gives $\lambda_\theta \approx 0.96$, a shift of $\sim 4\%$ easily resolvable with modern datasets~\cite{BESIII:2024lks}.

\begin{figure}[htbp!]
    \centering
    \includegraphics[width=0.9\linewidth]{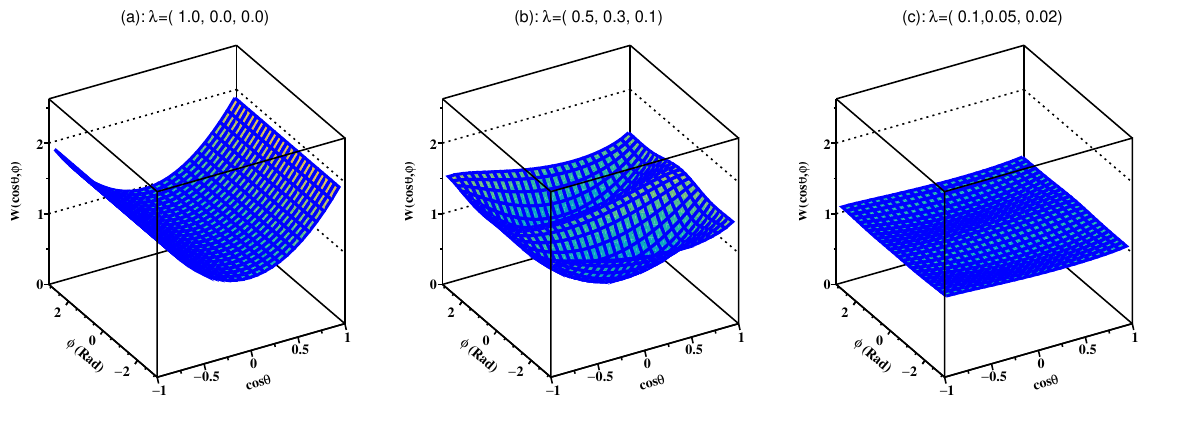}
    \caption{Two-dimensional dilepton angular distribution $W(\theta,\phi)$ from Eq.~\eqref{eq:WthetaPhi} for different polarization states. (a) Full transverse polarization: $\lambda_\theta=1, \lambda_\phi=\lambda_{\theta\phi}=0$. (b) Mixed polarization: $\lambda_\theta=0.5, \lambda_\phi=0.3, \lambda_{\theta\phi}=0.1$. (c) Nearly isotropic: $\lambda_\theta=0.1, \lambda_\phi=0.05, \lambda_{\theta\phi}=0.02$. The distributions demonstrate how SDM elements shape experimentally observable patterns.\label{fig:2d-angular-pol}}
\end{figure}

Figure~\ref{fig:2d-angular-pol} illustrates how different polarization configurations manifest in the $(\cos\theta, \phi)$ plane. Panel (a) shows the characteristic pattern for full transverse polarization ($\lambda_\theta = 1$), while panels (b) and (c) demonstrate progressive isotropization as polarization decreases.

\subsubsection{Statistical sensitivity via Fisher information}

\begin{figure}[htbp!]
    \centering
    \includegraphics[width=0.32\linewidth]{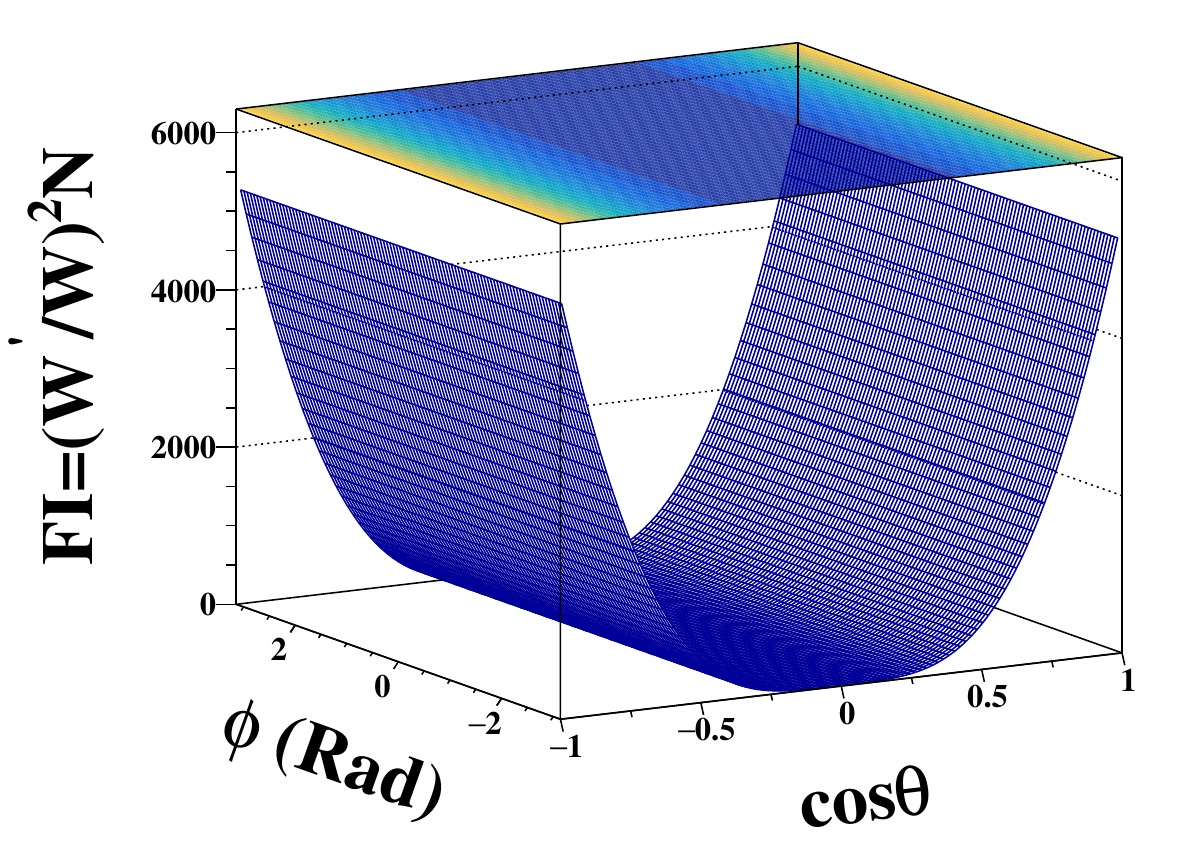}
    \includegraphics[width=0.32\linewidth]{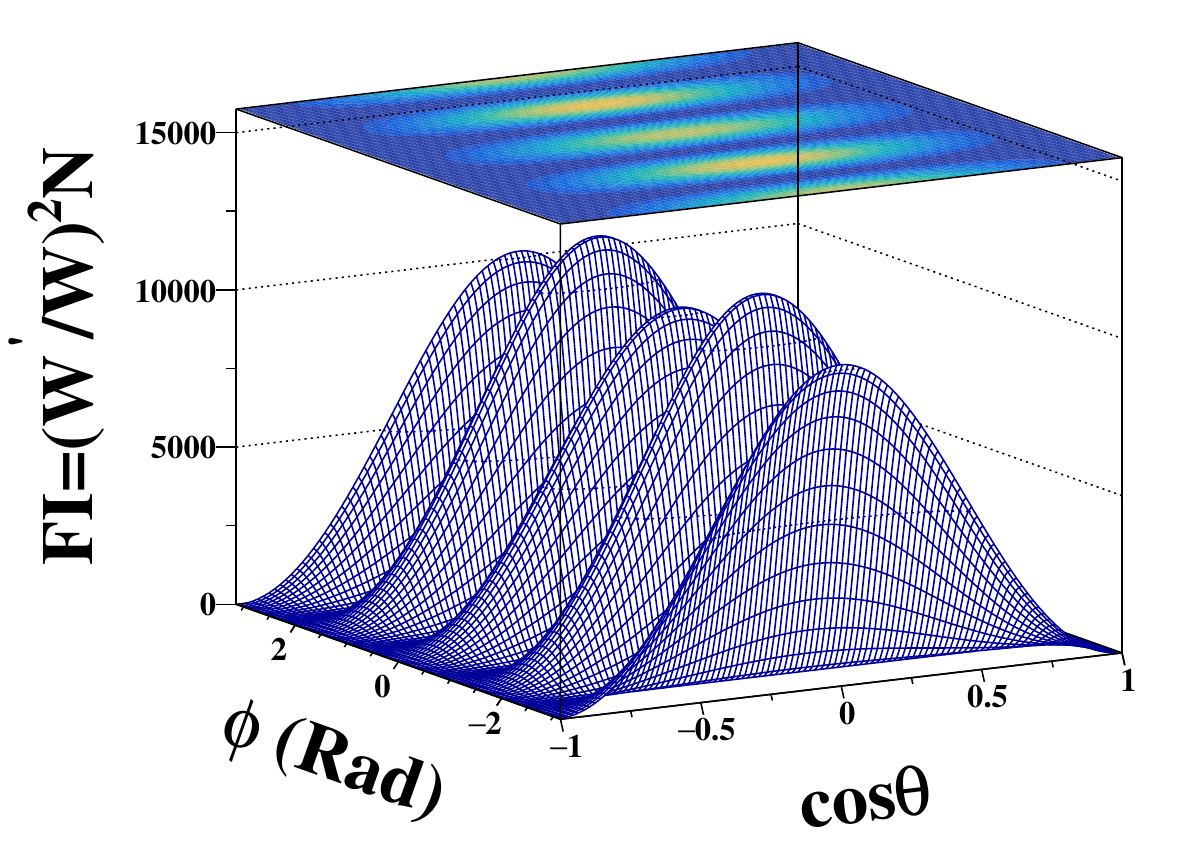}
    \includegraphics[width=0.32\linewidth]{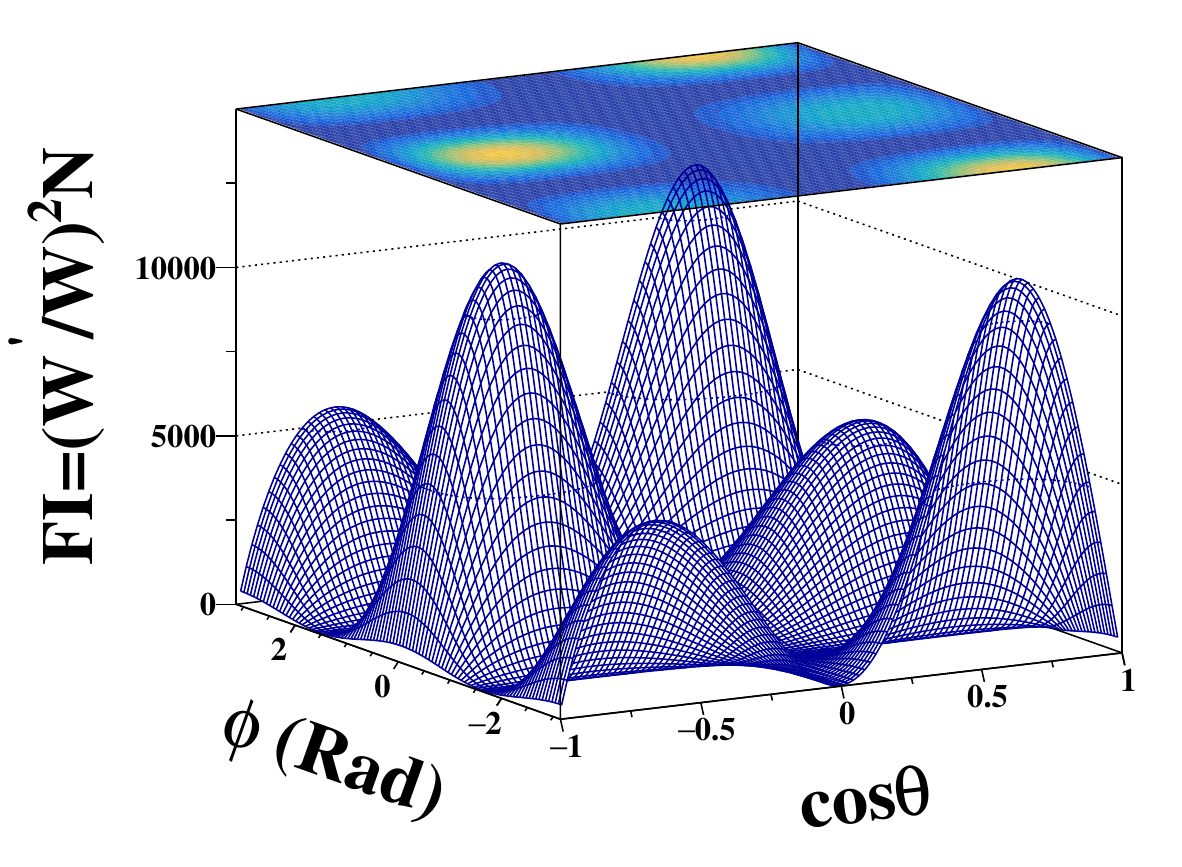}
    \caption{Observed Fisher information for (left) $\lambda_\theta$, (middle) $\lambda_\phi$, and (right) $\lambda_{\theta\phi}$ on the $(\cos\theta, \phi)$ plane. Higher Fisher information (bright regions) indicates phase space regions that provide more statistical sensitivity. For $\lambda_\theta$, sensitivity peaks near the beam axis ($|\cos\theta| \to 1$). For $\lambda_\phi$ and $\lambda_{\theta\phi}$, maximum sensitivity occurs at intermediate polar angles with specific azimuthal structure.\label{fig:fi-2d-angular}}
\end{figure}

The statistical precision with which polarization parameters can be extracted is quantified by the Fisher information~\cite{brandt2014data}:
\begin{equation}
\mathcal{I}_\alpha(\theta_i,\phi_i) = N_i\mathbb{E} \left[ \frac{\frac{\partial W(\theta_i,\phi_i)}{\partial \alpha}} {W(\theta_i,\phi_i)}\right]^2,
\end{equation}
where $\alpha \in \{\lambda_\theta, \lambda_\phi, \lambda_{\theta\phi}\}$, and $N_i$ and $\theta_i (\phi_i)$ are the number of events and the polar (azimuthal) angle in the $i$th grid, respectively. Fig.~\ref{fig:fi-2d-angular} shows the observed Fisher information distributions in the $(\cos\theta, \phi)$ plane for each parameter.

Integrating the observed Fisher information over the $(\cos\theta, \phi)$, adopting the BESIII's total number of $\psi^\prime$ events~\cite{BESIII:2024lks} $(2.712\pm 0.014)\times 10^9$ and a 40\% detection efficiency, the projected statistical uncertainties can be derived:

\begin{equation}
\delta\lambda_\theta \approx 0.0003, \qquad \delta\lambda_\phi \approx 0.0002, \qquad \delta\lambda_{\theta\phi} \approx 0.0002.
\end{equation}
This exceptional precision makes the method sensitive to $D$-wave fractions at the 1\% level.

\subsection{Self-consistent test of the polarization transfer framework}

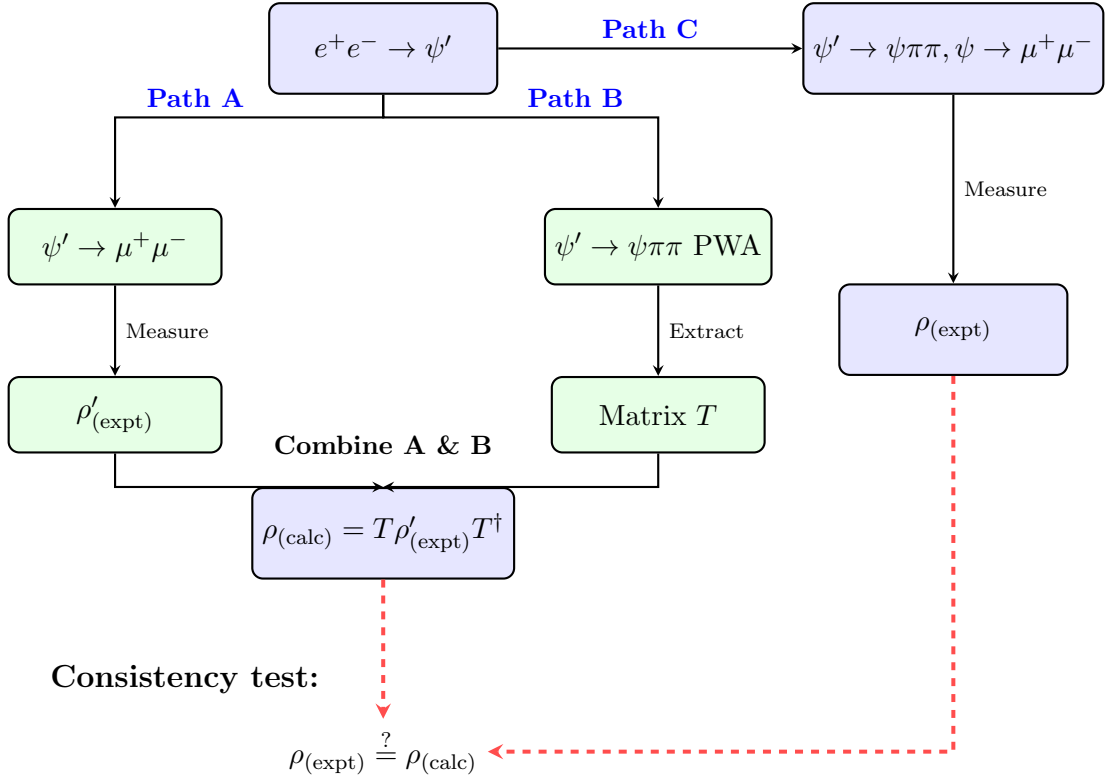
\begin{figure}[htbp]
\centering
\begin{tikzpicture}[
    node distance=1.8cm and 1.2cm,
    box/.style={draw, rectangle, minimum width=3.0cm, minimum height=1.2cm, 
                text centered, rounded corners, fill=blue!10, thick},
    smallbox/.style={draw, rectangle, minimum width=2.8cm, minimum height=1.0cm, 
                     text centered, rounded corners, fill=green!10, thick},
    arrow/.style={->, >=stealth, thick, black},
    dashedarrow/.style={->, >=stealth, thick, dashed, red!70}
]

\node[box] (production) {$e^+e^- \to \psi'$};

\node[smallbox, below left=1.5cm and 0.6cm of production] (psi2decay) 
    {$\psi' \to \mu^+\mu^-$};
\node[smallbox, below=1.2cm of psi2decay] (rhoprime) 
    {$\rho'_{\text{(expt)}}$};

\node[smallbox, below right=1.5cm and 0.6cm of production] (transition) 
    {$\psi' \to \psi\pi\pi$ PWA};
\node[smallbox, below=1.2cm of transition] (Tmatrix) 
    {Matrix $T$};

\node[box, below=5.2cm of production] (rhocalc) 
    {$\rho_{\text{(calc)}} = T\rho'_{\text{(expt)}}T^\dagger$};

\node[box, right=4.0cm of production] (decaychain) 
    {$\psi' \to \psi\pi\pi, \psi \to \mu^+\mu^-$};
\node[box, below=2.5cm of decaychain] (rhoexpt) 
    {$\rho_{\text{(expt)}}$};

\node[below=1.0cm of rhocalc] (comparison) 
    {\large\textbf{Consistency test:~~~~~~~~~~~~~~~~~~~~~~~~~~~~~~~~~~}};
\node[below=0.2cm of comparison] (comparisoneq) 
    {$\rho_{\text{(expt)}} \stackrel{?}{=} \rho_{\text{(calc)}}$};

\draw[arrow] (production.south) -- ++(0,-0.3) -| 
    node[pos=0.35, above, sloped, font=\small\bfseries] {\textcolor{blue}{Path A}} 
    (psi2decay.north);
\draw[arrow] (production.south) -- ++(0,-0.3) -| 
    node[pos=0.35, above, sloped, font=\small\bfseries] {\textcolor{blue}{Path B}} 
    (transition.north);
\draw[arrow] (production.east) -- 
    node[above, pos=0.5, font=\small\bfseries] {\textcolor{blue}{Path C}} 
    (decaychain.west);

\draw[arrow] (psi2decay) -- node[right, pos=0.5, font=\scriptsize] {Measure} (rhoprime);
\draw[arrow] (transition) -- node[right, pos=0.5, font=\scriptsize] {Extract} (Tmatrix);
\draw[arrow] (decaychain) -- node[right, pos=0.5, font=\scriptsize] {Measure} (rhoexpt);

\draw[arrow] (rhoprime.south) |- ([xshift=-0.8cm]rhocalc.north) -- (rhocalc.north); 
\draw[arrow] (Tmatrix.south) |- ([xshift=0.8cm]rhocalc.north) -- (rhocalc.north);
\node[above=0.3cm of rhocalc, font=\small\bfseries] {Combine A \& B};

\draw[dashedarrow, line width=1.5pt] (rhocalc.south) -- (comparisoneq);
\draw[dashedarrow, line width=1.5pt] (rhoexpt.south) |- 
    ([xshift=3.5cm]comparisoneq.west) -- (comparisoneq.east);

\end{tikzpicture}
\caption{Three-path self-consistent test of polarization transfer. Path A measures the parent SDM $\rho'_{\text{(expt)}}$ from $\psi' \to \mu^+\mu^-$. Path B extracts the transition matrix $T$ from PWA of $\psi' \to \psi\pi\pi$. These are combined to calculate $\rho_{\text{(calc)}}$. Path C directly measures the daughter SDM $\rho_{\text{(expt)}}$ from $\psi \to \mu^+\mu^-$ in the decay chain. Consistency between $\rho_{\text{(calc)}}$ and $\rho_{\text{(expt)}}$ validates the entire framework and constrains partial wave amplitudes.\label{fig:self-consistent-test}}
\end{figure}

Combining three independent measurements provides a powerful validation of the entire formalism, illustrated in Fig.~\ref{fig:self-consistent-test}. The first measurement (Path A) determines $\rho^\prime_{\text{(expt)}}$ by fitting the angular distribution of $\psi^\prime \to \mu^+\mu^-$ to Eq.~\eqref{eq:WthetaPhi}. This establishes the baseline polarization state of the directly produced $\psi^\prime$. The second measurement (Path B) involves an independent PWA of the $\psi$ and $\pi\pi$ angular distributions in $\psi^\prime \to \psi\pi\pi$, which determines the amplitudes $M_{001}, M_{201}, M_{021}$ that define the transition matrix $T$ via Eq.~\eqref{eqn:amp-full}. The third measurement (Path C) directly extracts the $\psi$ SDM from the dilepton angular distribution of $\psi \to \ell^+\ell^-$ within the decay chain $\psi^\prime \to \psi\pi\pi$, yielding $\rho_{\text{(expt)}}$ through the same fitting procedure.

This test proceeds in two stages. First, using the inputs from Paths A and B, the expected daughter SDM is calculated as $\rho_{\text{(calc)}} = T\rho^\prime_{\text{(expt)}}T^\dagger$. Second, the prediction is compared with the directly measured $\rho_{\text{(expt)}}$ from Path C. Statistical agreement validates the entire chain: the initial $\psi^\prime$ polarization measurement, the extracted partial wave amplitudes, and the SDM transformation formalism itself. Significant deviation would point to either residual systematic effects or missing dynamical components in the transition amplitude.

The exceptional precision outlined in Sec.~4.1 makes this test highly constraining. 
The expected $\delta\lambda_\theta \approx 0.0003$ provides stringent tests of QCD multipole predictions.
This methodology thereby establishes $\psi^\prime \rightarrow \psi\pi\pi$ as a benchmarked process for polarization calibration in quarkonium physics, with the over-determined structure of the test offering unprecedented systematic control.
\noindent
Although Fig.~\ref{fig:self-consistent-test} illustrates three conceptually independent paths, the optimal experimental strategy is a combined fit to the full angular information $(\Omega_\psi, \Omega_\pi, \Omega_\ell)$. This single fit simultaneously determines the partial-wave amplitudes $M_{\ell LS}$ and the $\psi$ SDM, exploiting the exact mathematical relation $\rho = T\rho' T^\dagger$ as an internal consistency constraint. Such a unified treatment yields the highest statistical power and provides the most rigorous test of the polarization-transfer framework.

\subsection{\texorpdfstring
  {Extension to $\psi^\prime \rightarrow h_c\pi^0$}
  {Extension to psi' to hc pi0}}
  
\begin{figure}[htbp]
\centering 
\begin{tikzpicture}[line width=0.9pt, font=\small, scale=1.0]
    \draw (-3, 0.0) -- (3, 0.0);
    \draw (-3,-0.5) -- (3,-0.5);
    
    \node[below] at (-2,-0.6) {$c$};
    \node[above] at (-2, 0.0) {$\bar{c}$};
    \node[below] at ( 2,-0.6) {$c$};
    \node[above] at ( 2, 0.0) {$\bar{c}$};
    
    \draw[decorate, decoration={snake, segment length=3mm, amplitude=1mm}, black!70] 
         (-0.2, 0.20) -- (-0.2,1.30);
    \draw[decorate, decoration={snake, segment length=3mm, amplitude=1mm}, black!70] 
         ( 0.2, 0.20) -- ( 0.2,1.30);
    
    \fill (0, 1.75) circle (0.50);
    \fill (0,-0.25) circle (0.50);
    \draw[dashed] (0.0,1.75) -- (3.0,1.75) node[right] {$\pi^0$};
    \node[right] at (3.6,1.75) {$j=0$};
    
    \node[left] at (-3.2,-0.2) {$\psi'$: $m=\pm1$};
    \node[right] at (3.0,-0.2) {$h_c$: $s_z=\pm1$};
    \node[right] at (3.6,0.85) {$L=0,2$};
\end{tikzpicture}
\caption{Schematic for $\psi' \to h_c\pi^0$. Only $L=0,2$ orbital angular momenta are allowed.\label{fig:pi0hc-diagram}}
\end{figure}
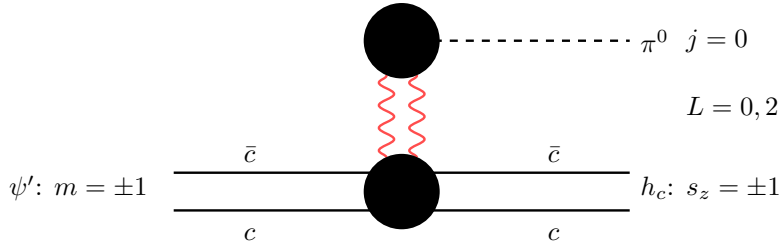

This formalism extends naturally to other hadronic transitions. Consider:
\begin{equation}\label{eq:hc-process}
e^+e^- \to \psi' \to h_c\pi^0,
\end{equation}
where $h_c$ ($J^{PC} = 1^{+-}$) is the $^1P_1$ charmonium state~\cite{Armstrong:1992ae, CLEO:2005vqq, BESIII:2010gid}. Since $\pi^0$ is a pseudoscalar, the transition is governed solely by the orbital angular momentum $L$ between $\pi^0$ and $h_c$, with allowed values $L = 0, 2$ (Fig.~\ref{fig:pi0hc-diagram}).

The amplitude is:
\begin{equation}
T_{s_z,m}(\Omega_{h_c}) = \sum_{L,L_z} M_L Y_L^{L_z}(\Omega_{h_c}) C^{1;Ls}_{m;L_z s_z} = M_0 \delta_{m s_z} + M_2\sum_{L_z} Y_L^{L_z}(\Omega_{h_c}) C^{1;Ls}_{m;L_z s_z}.
\end{equation}

For pure $S$-wave ($M_2 = 0$),  $\rho_{h_c} = \rho_{\psi'}$ is obtained again, demonstrating that polarization inheritance is a generic feature of $S$-wave soft-pion transitions. The $D$-wave corrections follow the same structure as Eqs.~\eqref{eq:rhojpsi-diag1}--\eqref{eq:rhojpsi-diag3}.

This transition provides a complementary probe: $\psi' \to \psi\pi\pi$ involves an $S$-wave to $S$-wave transition, while $\psi' \to h_c\pi^0$ involves $S$-wave to $P$-wave. Together, these channels map out polarization dynamics across charmonium multiplets.

\section{Discussion and conclusions\label{sec:conclude}}

A spin density matrix framework is developed for polarization transfer in $\psi^\prime \to \psi\pi\pi$, deriving the explicit transformation $\rho = T\rho^\prime T^\dagger$ that connects the polarization states of parent and daughter mesons. The central result is that for pure $S$-wave $\pi\pi$ emission --- the dominant $E1E1$ mechanism --- the SDM is perfectly preserved: $\rho = \rho^\prime$. This identity is a direct consequence of angular momentum conservation and provides an experimentally testable signature of the underlying dynamics. Deviations from this identity arise from $D$-wave contributions, which have been quantified analytically, demonstrating that even modest $D$-wave fractions of a few percent produce measurable shifts in the $\psi$ polarization parameters. The method thus converts a precision measurement of the $\psi$ SDM into a clean, model-independent constraint on sub-leading partial wave amplitudes.

Using the large BESIII $\psi^\prime$ data sample~\cite{BESIII:2024lks} with projected statistical uncertainties of $\delta\lambda_\theta \sim 0.0003$, the polarization-transfer technique can constrain $D$-wave fractions to about 1\%. The three-path self-consistent test illustrated in Fig.~\ref{fig:self-consistent-test} provides an over-constrained validation of the entire chain --- parent SDM, transition amplitudes, and daughter SDM --- offering improved systematic control.

Key applications involve two interrelated aspects: amplitude analyses and PWA of $\psi$ decays. Conventional measurements using $e^+e^-$ scan data at BESIII~\cite{BESIII:2025uin,BESIII:2018wid} are complicated by the interference of multiple amplitudes ---typically the $\psi\to ggg$ strong decay, the $\psi\to\gamma^*$ electromagnetic decay, and the non-resonant $e^+e^-$ continuum. The presence of two or more coherent amplitudes inevitably leads to multiple-solution ambiguities in the extracted magnitudes and phases. The transition $\psi^\prime \to \psi\pi\pi$ circumvents this difficulty: the $\psi$ is produced via a well-understood hadronic transition without continuum contamination, and its spin density matrix is  known. For amplitude analyses, this removes the dominant source of ambiguity, enabling clean extraction of $\psi$ decay amplitudes that would otherwise require model-dependent assumptions or extra data sets.
For PWA of $\psi$ decays, the same advantages apply. With a precisely known initial SDM and a continuum-free sample, the subsequent PWA proceeds as in conventional analyses but without the need to model interference with non-resonant continuum contributions.

The transformation law is intrinsic to the decay and independent of the production mechanism of the parent meson.  As long as the initial spin density matrix $\rho^\prime$ is known, the same transformation applies universally, rendering the approach portable across experimental environments. In $e^+e^-$ colliders like BESIII~\cite{BESIII:2009fln}, Belle II~\cite{Belle-II:2010dht}, or a future super-$\tau$-charm factory~\cite{Achasov:2023gey}, the $\psi^\prime$ is produced with a known polarization; at the LHC~\cite{Evans:2008zzb}, even unpolarized or weakly polarized production can provide consistency checks of $E1E1$ dominance. When $\psi^\prime$ is produced with a measurable non-trivial polarization, e.g.~in $B$-hadron decays or associated production, the same SDM transformation offers a unified probe across collider platforms.

The formalism extends directly to other quarkonium systems, e.g., $\Upsilon(nS) \to \Upsilon(mS)\pi\pi$, where the same $E1E1$ dominance and SDM preservation are expected. Comparative studies across charmonium and bottomonium can test the universality of non-perturbative QCD dynamics. The same angular-momentum structure appears in electroweak processes such as $e^+e^- \to Z^\ast \to ZH$ at future Higgs factories~\cite{3085641, FCC:2018evy}. The $S$-wave ($L=0$) contribution preserves the $Z$ polarization exactly, while $D$-wave ($L=2$) terms introduce mixing, providing an analogous test of the $HZZ$ coupling. These features establish polarization transfer as a versatile and robust probe of dynamics across energy regimes, from charmonium to the Higgs sector, with immediate applications at existing facilities and at future colliders.

\appendix
\section{\texorpdfstring
  {Explicit forms of transition matrices $T^{\ell LS}$}
  {Explicit forms of transition matrices TlLS}
}\label{sec:TlLS}

The transition matrices in the helicity basis $(s_z = +1, 0, -1) \times (m = +1, 0, -1)$ are:

\subsection{Pure $S$-wave: $(\ell,L,S) = (0,0,1)$}
\label{sec:T001}

\begin{equation}
T^{001}_{s_z m} = \frac{M_{001}}{4\pi} \delta_{s_z m}
= \frac{M_{001}}{4\pi}
\begin{pmatrix}
1 & 0 & 0 \\
0 & 0 & 0 \\
0 & 0 & 1
\end{pmatrix}~.
\end{equation}

\subsection{$D$-wave contributions: $(\ell,L,S) = (2,0,1)$ and $(0,2,1)$}
\label{sec:TDwaves}

The two $D$-wave amplitudes share an identical matrix structure, differing only in their angular arguments. For $(\ell,L,S)=(2,0,1)$ (dipion $D$-wave), the matrix depends on the dipion decay angles $(\theta_\pi,\phi_\pi)$; for $(\ell,L,S)=(0,2,1)$ (relative $D$-wave), it depends on the $\psi$ emission angles $(\theta_\psi,\phi_\psi)$. Both can be expressed as:

\begin{equation}
T^{D}_{s_z m}(\theta,\phi) =
\frac{\sqrt{2} M_{D}}{16\pi}
\begin{pmatrix}
3\cos^2\theta-1 & 0 & 3e^{-2i\phi}\sin^2\theta \\
\frac{3}{\sqrt{2}}e^{i\phi}\sin 2\theta & 0 & -\frac{3}{\sqrt{2}}e^{-i\phi}\sin 2\theta \\
3e^{2i\phi}\sin^2\theta & 0 & 3\cos^2\theta-1
\end{pmatrix}~,
\end{equation}
where $M_D$ represents the corresponding partial wave amplitude ($M_{201}$ or $M_{021}$), and $(\theta,\phi)$ are the appropriate angles: $(\theta_\pi,\phi_\pi)$ for $T^{201}$ and $(\theta_\psi,\phi_\psi)$ for $T^{021}$.

These explicit forms make the transformation $\rho = T\rho^\prime T^\dagger$ fully transparent for numerical implementation.

\acknowledgments
This work is supported by the National Key Research and Development Program of China under Grant No. 2025YFA1613900. The authors acknowledge the use of AI-assisted tools for language polishing and manuscript formatting. The authors thank Prof. Hu Zhen from Tsinghua University, and Profs. Li Yiming, Liao Hongbo, and Huang Yanping from the Institute of High Energy Physics, Chinese Academy of Sciences, for discussions on the generalization of the method.

\bibliographystyle{JHEP/JHEP}
\bibliography{JHEP/biblio}

\end{document}